\documentclass[aps,twocolumn,showpacs]{revtex4}
\usepackage{amsmath}
\usepackage{epsfig}

\begin{document}

\title{Investigating the hidden-charm and hidden-bottom pentaquark resonances in scattering process.}

\newcommand*{\NJNU}{Department of Physics, Nanjing Normal University, Nanjing, Jiangsu 210097, China}\affiliation{\NJNU}

\author{Hongxia Huang}\email{hxhuang@njnu.edu.cn}\affiliation{\NJNU}
\author{Jialun Ping}\email{jlping@njnu.edu.cn}\affiliation{\NJNU}

\begin{abstract}
In the framework of quark delocalization color screening model, both the hidden-charm and
hidden-bottom pentaquark resonances are studied in the hadron-hadron scattering process.
A few narrow pentaquark resonances with hidden-charm above $4.2$ GeV, and some narrow
pentaquark resonances with hidden-bottom above $11$ GeV are found from corresponding
scattering processes. Besides, the states $N\eta_{c}$, $NJ/\psi$, $N\eta_{b}$ and $N\Upsilon$
with $IJ^{P}=\frac{1}{2}\frac{1}{2}^{-}$, as well as $NJ/\psi$ and $N\Upsilon$ with
$IJ^{P}=\frac{1}{2}\frac{3}{2}^{-}$ are all possible to be bound by channel-coupling calculation.
All these heavy pentaquarks are worth searching in the future experiments.
\end{abstract}


\maketitle

\setcounter{totalnumber}{5}

\section{\label{sec:introduction}Introduction}

In 2003, the LEPS Collaboration announced the observation of the pentaquark $\Theta^{+}$~\cite{Nakano},
which inspired a lot of theoretical work, as well as experimental work to search for pentaquarks.
However, this state was not confirmed by the subsequent more advanced experiments.
Nevertheless, the LEPS Collaboration still insists on the existence of pentaquark
$\Theta^+$~\cite{NakanoNew}, and the relevant experiment is also in progress~\cite{Megumi}.
Moreover, there were also some theoretical studies on the existence of the hidden-charm pentaquarks~\cite{WuJJ,YangZC,XiaoCW1,Uchino,Karliner,Garzon,WangWL,YuanSG,HuangY}.
In the year of 2015, the claim of two hidden-charm pentaquark states $P_{c}(4380)$ and
$P_{c}(4450)$ by the LHCb Collaboration~\cite{LHCb} attracted people's interesting in
the pentaquarks again and triggered more and more theoretical work on these two states.
Until now, theoretical interpretations of $P_{c}(4380)$ and $P_{c}(4450)$ include the
baryon-meson molecules~\cite{ChenR1,ChenHX1,RocaL,HeJ,HuangHX,GangY,Meissner,XiaoCW2,ChenR2,ChenHX2},
the diquark-triquark pentaquarks~\cite{Lebed,ZhuR},
the diquark-diquark-antiquark pentaquarks~\cite{Maiani,Anisovich,Ghosh,WangZG},
the genuine multiquark states~\cite{Mironov}, the topological soliton~\cite{Scoccola},
and the kinematical threshold effects in the triangle singularity mechanism~\cite{GuoFK,LiuXH1,Mikhasenko},
etc. The more comprehensive discussions on the current experimental progresses
and various theoretical interpretations of these candidates can be found in Ref.~\cite{ChenHX3}.

%In fact, many multiquark states are resonance states, which are investigated in the related hadron-hadron scattering process in experiments. Resonances are %unstable particles usually observed as bell-shaped structures in scattering cross sections of their open channels. For a simple narrow resonance, its %fundamental properties correspond to the visible cross-section features: mass is at the peak position, and decay width is the half-width of the bell shape.
To provide the necessary information for experiments to search for multiquark states, mass spectrum
calculation alone is not enough. The calculation of hadron-hadron scattering, the main production
process of multiquark states, is indispensable. The scattering phase shifts will show a resonance
behavior in the resonance energy region. In many theoretical work mentioned above, they investigated
$P_{c}(4380)$ and $P_{c}(4450)$ as bound states. In fact, these states will decay through the related
open channels. As we mentioned in our previous work~\cite{HuangHX}, in the bound-state calculation
$P_{c}(4380)$ can be explained as the molecular pentaquark $\Sigma^{*}_{c}D$ with the quantum number
$J^{P}=\frac{3}{2}^{-}$, but it can decay to the open channels $NJ/\psi$ and $\Lambda_{c}D^{*}$.
Therefore, we should study the $NJ/\psi$ and $\Lambda_{c}D^{*}$ scattering process to check whether
the $P_{c}(4380)$ is a resonance state or not.
Similar work has been done in the dibaryon system, in which we obtained the $d^{*}$ resonance
during the $NN$ scattering process, and found that the energy and decay width of the partial wave
of $NN$ were consistent with the experiment data~\cite{Ping2009}. Extending to the pentaquark system,
we investigated the $N\phi$ state in the different scattering channels: $N\eta'$, $\Lambda K$, and
$\Sigma K$~\cite{Gao}. Both the resonance mass and decay width were obtained, which provided the
necessary information for experimental searching at Jefferson Lab. Therefore, it is interesting
to extend such study to the molecular pentaquarks with heavy quarks.

Generally, hadron structure and hadron interactions belong to the low energy physics of quantum
chromodynamics (QCD), which are much harder to calculate directly from QCD because of the non-perturbative
nature of QCD. One has to rely on effective theories and/or QCD-inspired models to get some insight
into the phenomena of the hadronic world. The constituent quark model is
one of them, which approximately transforms the complicated interactions between current quarks
into dynamic properties of quasiparticles (constituent quark) and considers the residual
interactions between quasiparticles. There are various kinds of constituent quark models,
such as one-boson-exchange model, chiral quark model, quark delocalization color screening model (QDCSM),
and so on. These models have been successful in describing hadron spectrum, the baryon-baryon
interactions and the bound state of two baryons, the deuteron. Among these phenomenological models,
QDCSM, which was developed in the 1990s with the aim of explaining the similarities between nuclear
(hadronic clusters of quarks) and molecular forces~\cite{QDCSM0}, was extensively used and studied
in our group. In this model, quarks confined in
one cluster are allowed to delocalize to a nearby cluster and the
confinement interaction between quarks in different clusters is
modified to include a color screening factor. The latter is a model
description of the hidden color channel coupling
effect~\cite{QDCSM1}. The delocalization parameter is determined by
the dynamics of the interacting multi-quark system, thus allows the system to choose the most
favorable configuration through its own dynamics in a larger Hilbert space. Recently, this model
has been used to study the hidden-charm pentaquarks~\cite{HuangHX}. We found that the interaction
between $\Sigma_{c}$ (or $\Sigma_{c}^{*}$) and $D$ (or $D^{*}$) was strong enough to form some
bound states, and $P_{c}(4380)$ can be interpreted as the molecular state $\Sigma^{*}_{c}D$ with
quantum numbers $IJ^{P}=\frac{1}{2}\frac{3}{2}^{-}$. In this work, we continue to study the
hidden-charm resonance states in the related hadron-hadron scattering process. Besides, we also
extend the study to the hidden bottom sector to search for the some hidden-bottom pentaquark resonances.

In the next section, the framework of the QDCSM and the calculation method are briefly
introduced. Section III devotes to the
numerical results and discussions. The summary is shown in the
last section.

\section{Quark model and the calculation method}

Since our previous work of the bound-state calculation of the hidden-charm molecular
pentaquark~\cite{HuangHX} was carried through our quark delocalization color screening model (QDCSM),
we use the same model and parameters to study the pentaquark resonances here.
Besides, to calculate the baryon-meson scattering phase shifts and
to observe the resonance states, the well developed resonating group
method (RGM)~\cite{RGM} is used.

\subsection{Quark delocalization color screening model}

The detail of QDCSM used in the present work can be found in the
references~\cite{QDCSM0,QDCSM1,QDCSM2}. Here, we
just present the salient features of the model. The model
Hamiltonian is:
\begin{widetext}
\begin{eqnarray}
H &=& \sum_{i=1}^5 \left(m_i+\frac{p_i^2}{2m_i}\right) -T_c
+\sum_{i<j} \left[ V^{G}(r_{ij})+V^{\chi}(r_{ij})+V^{C}(r_{ij})
\right],
 \nonumber \\
V^{G}(r_{ij})&=& \frac{1}{4}\alpha_{s} {\mathbf \lambda}_i \cdot
{\mathbf \lambda}_j
\left[\frac{1}{r_{ij}}-\frac{\pi}{2}\left(\frac{1}{m_{i}^{2}}+\frac{1}{m_{j}^{2}}+\frac{4{\mathbf
\sigma}_i\cdot {\mathbf\sigma}_j}{3m_{i}m_{j}}
 \right)
\delta(r_{ij})-\frac{3}{4m_{i}m_{j}r^3_{ij}}S_{ij}\right],
\nonumber \\
V^{\chi}(r_{ij})&=& \frac{1}{3}\alpha_{ch}
\frac{\Lambda^2}{\Lambda^2-m_{\chi}^2}m_\chi \left\{ \left[
Y(m_\chi r_{ij})- \frac{\Lambda^3}{m_{\chi}^3}Y(\Lambda r_{ij})
\right]
{\mathbf \sigma}_i \cdot{\mathbf \sigma}_j \right.\nonumber \\
&& \left. +\left[ H(m_\chi r_{ij})-\frac{\Lambda^3}{m_\chi^3}
H(\Lambda r_{ij})\right] S_{ij} \right\} {\mathbf F}_i \cdot
{\mathbf F}_j, ~~~\chi=\pi,K,\eta \\
V^{C}(r_{ij})&=& -a_c {\mathbf \lambda}_i \cdot {\mathbf
\lambda}_j [f(r_{ij})+V_0], \nonumber
\\
 f(r_{ij}) & = &  \left\{ \begin{array}{ll}
 r_{ij}^2 &
 \qquad \mbox{if }i,j\mbox{ occur in the same baryon orbit} \\
  \frac{1 - e^{-\mu_{ij} r_{ij}^2} }{\mu_{ij}} & \qquad
 \mbox{if }i,j\mbox{ occur in different baryon orbits} \\
 \end{array} \right.
\nonumber \\
S_{ij} & = &  \frac{{\mathbf (\sigma}_i \cdot {\mathbf r}_{ij})
({\mathbf \sigma}_j \cdot {\mathbf
r}_{ij})}{r_{ij}^2}-\frac{1}{3}~{\mathbf \sigma}_i \cdot {\mathbf
\sigma}_j. \nonumber
\end{eqnarray}
\end{widetext}
Where $S_{ij}$ is quark tensor operator; $Y(x)$ and $H(x)$ are
standard Yukawa functions~\cite{Valcarce}; $T_c$ is the kinetic
energy of the center of mass; $\alpha_{ch}$ is the chiral coupling
constant; determined as usual from the $\pi$-nucleon coupling
constant; $\alpha_{s}$ is the quark-gluon coupling constant. In
order to cover the wide energy range from light to heavy
quarks one introduces an effective scale-dependent quark-gluon
coupling $\alpha_{s}(\mu)$ \cite{Vijande}:
\begin{eqnarray}
\alpha_{s}(\mu) & = &
\frac{\alpha_{0}}{\ln(\frac{\mu^2+u_{0}^2}{\Lambda_{0}^2})}.
\end{eqnarray}
Where $\mu$ is the reduced mass of the interacting quarks pair. All other symbols have their usual
meanings, and all parameters are taken from our previous work~\cite{HuangHX}.
Besides, the quark delocalization in QDCSM is realized by specifying the
single particle orbital wave function of QDCSM as a linear
combination of left and right Gaussians, the single particle
orbital wave functions used in the ordinary quark cluster model. One can refer to the Ref.~\cite{HuangHX}
to see the orbital wave functions.

\subsection{The calculation method}
Here, we calculate the baryon-meson scattering phase shifts and
investigate the resonance states by using the resonating group method (RGM)~\cite{RGM}, a well
established method for studying a bound-state problem or a scattering one. The wave function of
the baryon-meson system is of the form
\begin{equation}
\Psi = {\cal A } \left[\hat{\phi}_{A}(\boldsymbol{\xi}_{1},\boldsymbol{\xi}_{2})
       \hat{\phi}_{B}(\boldsymbol{\xi}_{3})\chi_{L}(\boldsymbol{R}_{AB})\right].
\end{equation}
where $\boldsymbol{\xi}_{1}$ and $\boldsymbol{\xi}_{2}$ are the internal coordinates for the baryon
cluster A, and $\boldsymbol{\xi}_{3}$ is the internal coordinate for the meson cluster B.
$\boldsymbol{R}_{AB} = \boldsymbol{R}_{A}-\boldsymbol{R}_{B}$ is the relative coordinate between
the two clusters. The $\hat{\phi}_{A}$ and $\hat{\phi}_{B}$ are the internal cluster wave functions of
the baryon A (antisymmetrized) and meson B, and $\chi_{L}(\boldsymbol{R}_{AB})$ is the relative motion
wave function between two clusters. The symbol ${\cal A }$ is the anti-symmetrization operator defined as
\begin{equation}
{\cal A } = 1-P_{14}-P_{24}-P_{34},
\end{equation}
where 1, 2, and 3 stand for the quarks in the baryon cluster and 4 stands for the quark in the meson cluster.
For a bound-state problem, $\chi_{L}(\boldsymbol{R}_{AB})$ is expanded by gaussian bases
\begin{eqnarray}
& & \chi_{L}(\boldsymbol{R}_{AB}) = \frac{1}{\sqrt{4\pi}}(\frac{6}{5\pi b^2})^{3/4} \sum_{i=1}^{n} C_{i}  \nonumber \\
&& ~~~~\times  \int \exp\left[-\frac{3}{5b^2}(\boldsymbol{R}_{AB}-\boldsymbol{S}_{i})^{2}\right] Y_{LM}(\hat{\boldsymbol{S}_{i}})d\hat{\boldsymbol{S}_{i}} \nonumber \\
&& ~~~~~~~~~~~~= \sum_{i=1}^{n} C_{i} \frac{u_{L}(R_{AB},S_{i})}{R_{AB}}Y_{LM}(\hat{\boldsymbol{R}}_{AB}).
~~~~~
\end{eqnarray}
with
\begin{eqnarray}
& & u_{L}(R_{AB},S_{i}) = \sqrt{4\pi}(\frac{6}{5\pi b^2})^{3/4}R_{AB}   \nonumber \\
&& ~~~~\times \exp\left[-\frac{3}{5b^2}(R^{2}_{AB}-S^{2}_{i})\right] i^{L} j_{L}(-i \frac{6}{5b^{2}}R_{AB}S_{i}).
~~~~~
\end{eqnarray}
where $\boldsymbol{S}_{i}$ is called the generating coordinate, $C_{i}$ is expansion coefficients,
$n$ is the number of the gaussian bases, which is determined by the stability of the results,
and $j_{L}$ is the $L$-th spherical Bessel function.

For a scattering problem, the relative wave function is expanded as
\begin{equation}
\chi_{L}(\boldsymbol{R}_{AB}) = \sum_{i=1}^{n} C_{i}
    \frac{\tilde{u}_{L}(R_{AB},S_{i})}{R_{AB}}Y_{LM}(\hat{\boldsymbol{R}}_{AB}) .
\end{equation}
with
\begin{eqnarray}
 & & \tilde{u}_{L}(R_{AB},S_{i}) =    \nonumber \\
 & & \left\{ \begin{array}{ll}
 \alpha_{i} u_{L}(R_{AB},S_{i}),   ~~~~~~~~~~~~~~~~~~~~~~~~~~~~~~~~~~R_{AB}\leq R_{C}  \\
  \left[h^{-}_{L}(k_{AB},R_{AB})-s_{i}h^{+}_{L}(k_{AB},R_{AB})\right] R_{AB},  R_{AB}\geq R_{C}
 \end{array} \right. \nonumber \\
\end{eqnarray}
where $h^{\pm}_{L}$ is the $L$-th spherical Hankel functions, $k_{AB}$ is the momentum of relative motion
with $k_{AB}=\sqrt{2\mu_{AB}E_{cm}}$, $\mu_{AB}$ is the reduced mass of two hadrons (A and B) of the open
channel; $E_{cm}$ is the incident energy, and $R_{C}$ is a cutoff radius beyond which all the strong
interaction can be disregarded. Besides, $\alpha_{i}$ and $s_{i}$ are complex parameters which are
determined by the smoothness condition at $R_{AB}=R_{C}$ and $C_{i}$ satisfy $\sum_{i=1}^{n} C_{i}=1$.
After performing variational procedure, a $L$-th partial-wave equation for the scattering problem can be
deduced as
\begin{eqnarray}
& & \sum_{j=1}^{n} {\cal L }^{L}_{ij} C_{j} = {\cal M }^{L}_{i}  ~~~~(i=0,1,\cdot\cdot\cdot, n-1), \label{Lij}
~~~~~
\end{eqnarray}
with
\begin{eqnarray}
& & {\cal L }^{L}_{ij} = {\cal K }^{L}_{ij}-{\cal K }^{L}_{i0}-{\cal K }^{L}_{0j}+{\cal K }^{L}_{00},
~~~~~
\end{eqnarray}
\begin{eqnarray}
& & {\cal M }^{L}_{i}  = {\cal K }^{L}_{00}-{\cal K }^{L}_{i0},
~~~~~
\end{eqnarray}
and
%\begin{widetext}
\begin{eqnarray}
& & {\cal K }^{L}_{ij}=\left\langle \hat{\phi}_{A}(\boldsymbol{\xi}'_{1},\boldsymbol{\xi}'_{2})\hat{\phi}_{B}(\boldsymbol{\xi}'_{3})
  \frac{\tilde{u}_{L}(R'_{AB},S_{i})}{R'_{AB}}Y_{LM}(\hat{\boldsymbol{R}}'_{AB}) \right.  \nonumber \\
& & ~~~~~~~~~\left|H-E\right| \nonumber \\
& & \left. {\cal A } \left[\hat{\phi}_{A}(\boldsymbol{\xi}_{1},\boldsymbol{\xi}_{2})\hat{\phi}_{B}(\boldsymbol{\xi}_{3})
  \frac{\tilde{u}_{L}(R_{AB},S_{j})}{R_{AB}}Y_{LM}(\hat{\boldsymbol{R}}_{AB})\right]\right\rangle. \nonumber \\
\end{eqnarray}
%\end{widetext}
By solving Eq.(\ref{Lij}), we can obtain the expansion coefficients $C_{i}$. Then the $S$ matrix element $S_{L}$ and the phase shifts $\delta_{L}$ are given by
\begin{eqnarray}
& & S_{L} \equiv e^{2i\delta_{L}} = \sum_{i=1}^{n} C_{i}s_{i},
~~~~~
\end{eqnarray}

\section{The results and discussions}

\begin{figure*}
\begin{center}
\epsfxsize=5.5in \epsfbox{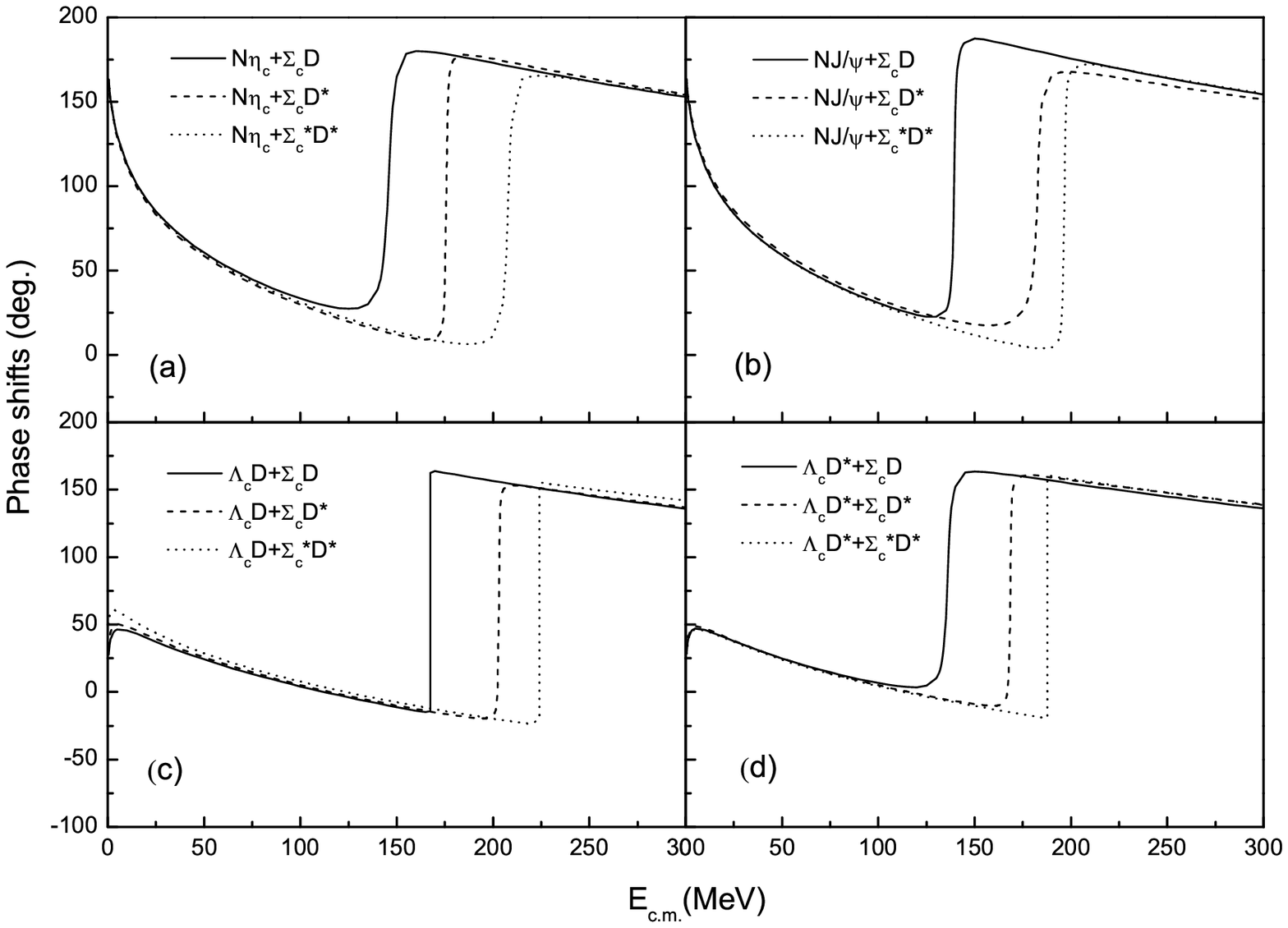} \vspace{-0.1in}

\caption{The $N\eta_{c}$, $NJ/\psi$, $\Lambda_{c}D$ and $\Lambda_{c}D^{*}$ $S-$wave phase shifts with two-channel coupling for the $IJ^{P}=\frac{1}{2}\frac{1}{2}^{-}$ system.}
\end{center}
\end{figure*}

\begin{table*}
\begin{center}
\caption{The mass and decay width (in MeV) of the $IJ^{P}=\frac{1}{2}\frac{1}{2}^{-}$ resonance states in the $N\eta_{c}$, $NJ/\psi$, $\Lambda_{c}D$ and $\Lambda_{c}D^{*}$ $S-$wave scattering process.}
{\begin{tabular}{@{}c|cccccc|cccccc} \hline
 \multicolumn{1}{c|}{}
 &\multicolumn{6}{c}{two-channel coupling}  & \multicolumn{6}{|c}{four-channel coupling}  \\\hline
\multicolumn{1}{c|}{}
 &\multicolumn{2}{c}{$\Sigma_{c} D$}&\multicolumn{2}{c}{$\Sigma_{c} D^{*}$}&\multicolumn{2}{c|}{$\Sigma_{c}^{*} D^{*}$} &\multicolumn{2}{c}{$\Sigma_{c} D$}&\multicolumn{2}{c}{$\Sigma_{c} D^{*}$}&\multicolumn{2}{c}{$\Sigma_{c}^{*} D^{*}$}\\
 &~~~~~$M$~~~ & ~~~$\Gamma$~~~ & ~~~$M$~~~ & ~~~$\Gamma$~~~ & ~~~$M$~~~ & ~~~$\Gamma$~~~  & ~~~~~$M$~~~ & ~~~$\Gamma$~~~ & ~~~$M$~~~ & ~~~$\Gamma$~~~ & ~~~$M$~~~ & ~~~$\Gamma$~~~ \\ \hline
 ~~~{$N\eta_{c}$}~~~        & ~~4309.8 & 6.0 & 4451.7 & 1.1 & 4523.1 & 3.5 & ~~4311.3 & 4.5 & 4448.8 & 1.0 & 4525.8 & 4.0  \\
 {$NJ/\psi$}          & ~~4305.9 & 2.0 & 4461.6 & 4.0 & 4514.7 & 1.2 & ~~4307.9 & 1.2 & 4459.7 & 3.9 & nr     & --   \\
 {$\Lambda_{c}D$}     & ~~4308.4 & 0.003 & 4452.6 & 1.0 & 4512.6 & 0.004 & ~~4306.7 & 0.02 & 4461.6 & 1.0 & nr & -- \\
 {$\Lambda_{c}D^{*}$} & ~~4311.6 & 3.5 & 4452.5 & 1.0 & 4510.8 & 0.005 & ~~4307.7 & 1.4 & 4449.0 & 0.3 & nr & --  \\
  \hline
\end{tabular}
\label{mass_c1}}
\end{center}
\end{table*}
From our previous bound-state calculation~\cite{HuangHX}, for the $IJ^{P}=\frac{1}{2}\frac{1}{2}^{-}$ system,
the single channel $\Sigma_{c}D$, $\Sigma_{c}D^{*}$ and $\Sigma^{*}_{c}D^{*}$ was bound; while the other four
channels $N \eta_{c}$, $N J/\psi$, $\Lambda_{c}D$ and $\Lambda_{c}D^{*}$ were unbound and scattering channels.
For the $IJ^{P}=\frac{1}{2}\frac{3}{2}^{-}$ system, the case is similar. There are two scattering channels
($N J/\psi$ and $\Lambda_{c}D^{*}$) and three bound-state channels ($\Sigma_{c}D^{*}$, $\Sigma^{*}_{c}D$ and
$\Sigma^{*}_{c}D^{*}$). These bound states may appear as resonance states in the corresponding scattering
channels and acquire finite widths. We should mention that all states we study here are in $S-$wave because
we found that there was no bound state with higher partial waves in our calculations. The $S-$wave bound
states decay to $D-$wave open channels through tensor interaction are neglected here due to the small decay widths.
So the total decay width of the states given below is the lower limits, also due to only the hidden-charm channels
are considered in this work. Besides, we do two kinds of channel-coupling in this work. The first one is the
two-channel coupling with a single bound state and a related open channel; another one is the four-channel coupling
with three bound states and a corresponding open channel. The general features of the calculated results are as follows.
\begin{figure}
\begin{center}
\epsfxsize=3.3in \epsfbox{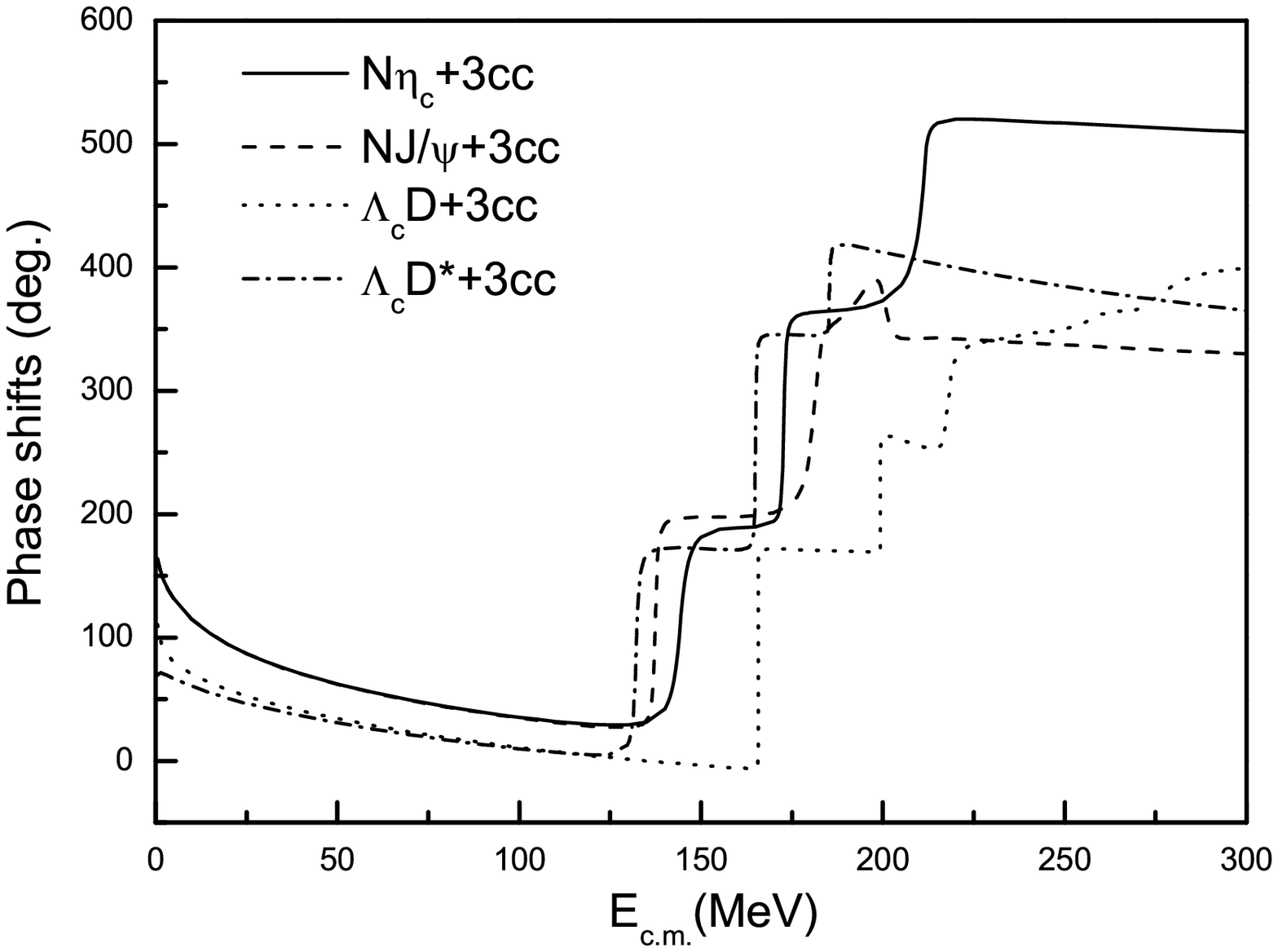} \vspace{-0.1in}

\caption{The $N\eta_{c}$, $NJ/\psi$, $\Lambda_{c}D$ and $\Lambda_{c}D^{*}$ $S-$wave phase shifts with
  four-channel coupling for the $IJ^{P}=\frac{1}{2}\frac{1}{2}^{-}$ system.}

\epsfxsize=3.3in \epsfbox{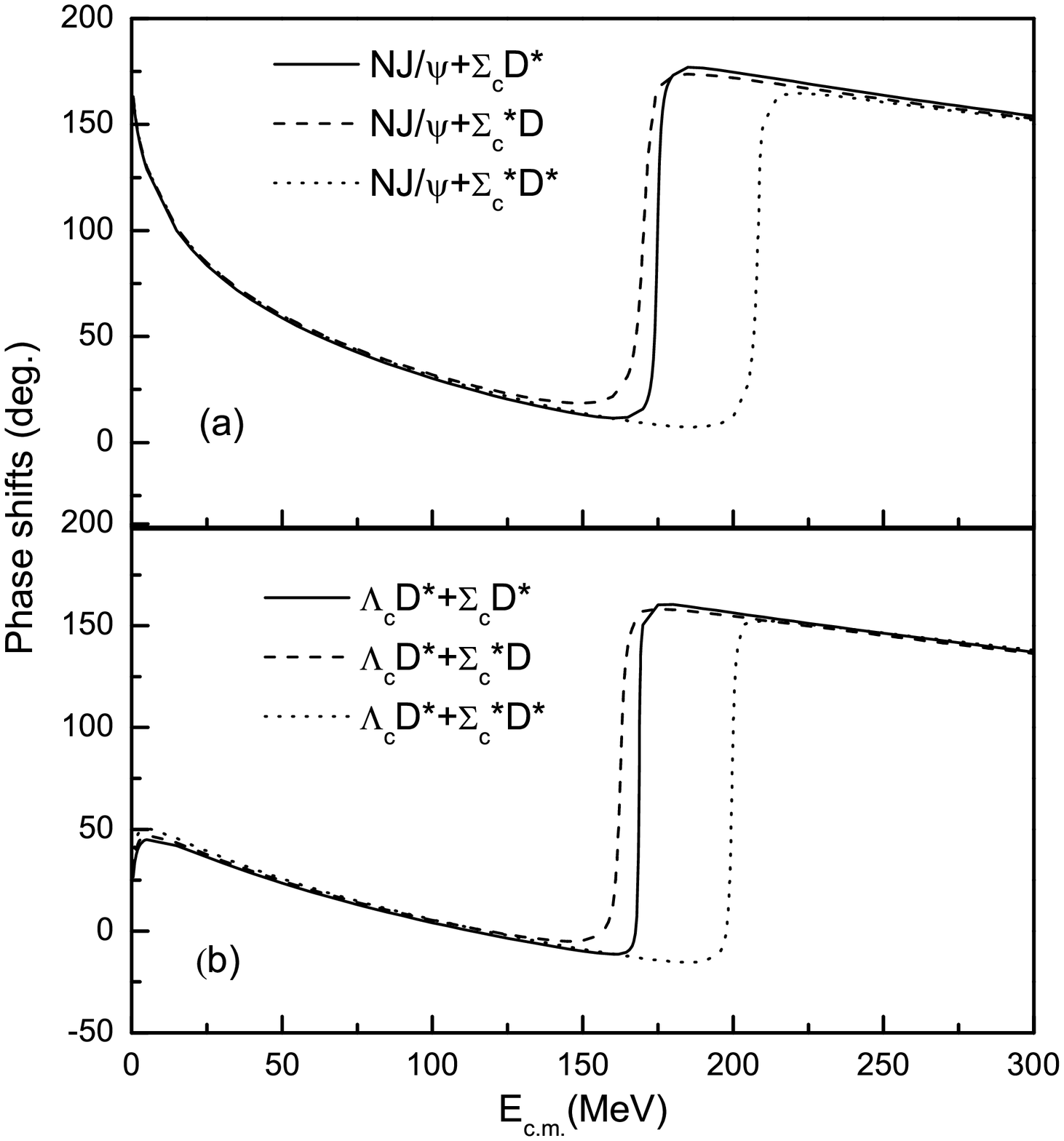} \vspace{-0.1in}

\caption{The $NJ/\psi$ and $\Lambda_{c}D^{*}$ $S-$wave phase shifts with two-channel coupling for
  the $IJ^{P}=\frac{1}{2}\frac{3}{2}^{-}$ system.}
\end{center}
\end{figure}

For the $IJ^{P}=\frac{1}{2}\frac{1}{2}^{-}$ system, we do the two-channel coupling calculation firstly.
The phase shifts of all scattering channels are shown in Fig. 1. The phase shifts of the $N\eta_{c}$ channel
(see Fig. 1(a)) clearly show three resonance states, which means that every bound state $\Sigma_{c}D$,
$\Sigma_{c}D^{*}$ and $\Sigma^{*}_{c}D^{*}$ appear as resonance state by coupling to the scattering channel
$N\eta_{c}$. Other scattering channels $NJ/\psi$, $\Lambda_{c}D$ and $\Lambda_{c}D^{*}$
(see Fig. 1 (b), (c) and (d)) show similar results as that of $N\eta_{c}$. From the shape of the resonance,
the resonance mass and decay width of every resonance state can be obtained, which are listed in Table~\ref{mass_c1}.
Comparing with the result of our previous bound-state calculation~\cite{HuangHX}, the mass shift of every resonance
state is not very large, which indicates that the scattering channel and the bound-state channel coupling effect
is not very strong although it is through the central force. The reasons is that the mass difference between the
scattering channel and the bound-state channel is large, which is about $100 \sim 400$ MeV.
\begin{table}
\begin{center}
\caption{The mass and decay width (in MeV) of the $IJ^{P}=\frac{1}{2}\frac{3}{2}^{-}$ resonance states
  in the $NJ/\psi$ and $\Lambda_{c}D^{*}$ $S-$wave scattering process.}
{\begin{tabular}{c|cccccc} \hline
  & \multicolumn{6}{c}{two-channel coupling}  \\ \hline
  & \multicolumn{2}{c}{$\Sigma_{c} D^{*}$} & \multicolumn{2}{c}{$\Sigma^{*}_{c} D$} & \multicolumn{2}{c}{$\Sigma_{c}^{*} D^{*}$} \\ &~~~~~$M$~~~ & ~~~$\Gamma$~~~ & ~~~$M$~~~ & ~~~$\Gamma$~~~ & ~~~$M$~~~ & ~~~$\Gamma$~~~  \\ \hline
~~{$NJ/\psi$}~~ & ~~4453.8 & 1.7 & 4379.7 & 4.5 & 4526.4 & 2.5  \\
 {$\Lambda_{c}D^{*}$} & ~~4452.7 & 0.8 & 4377.6 & 3.2 & 4522.7 & 1.8   \\   \hline
 & \multicolumn{6}{|c}{four-channel coupling}  \\ \hline
 & \multicolumn{2}{c}{$\Sigma_{c} D^{*}$} & \multicolumn{2}{c}{$\Sigma^{*}_{c} D$} & \multicolumn{2}{c}{$\Sigma_{c}^{*} D^{*}$} \\
 & ~~~~~$M$~~~ & ~~~$\Gamma$~~~ & ~~~$M$~~~ & ~~~$\Gamma$~~~ & ~~~$M$~~~ & ~~~$\Gamma$~~~ \\ \hline
 ~~{$NJ/\psi$}~~ & ~~4454.0 & 1.5 & 4376.4 & 1.5 & nr & --  \\
 {$\Lambda_{c}D^{*}$} & ~~4452.0 & 0.3 & 4374.4 & 0.9 & 4523.0 & 1.0  \\
  \hline
\end{tabular}
\label{mass_c2}}
\end{center}
\end{table}
\begin{figure}
\begin{center}
\epsfxsize=3.3in \epsfbox{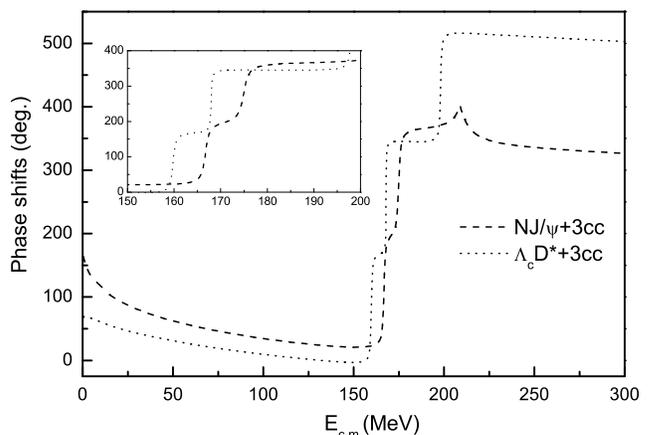} \vspace{-0.2in}

\caption{The $NJ/\psi$ and $\Lambda_{c}D^{*}$ $S-$wave phase shifts with four-channel coupling
  for the $IJ^{P}=\frac{1}{2}\frac{3}{2}^{-}$ system.}
\end{center}
\end{figure}

To investigate the effect of channel-coupling of the bound states, we also do the four-channel coupling
calculation. The phase shifts of all scattering channels of the $IJ^{P}=\frac{1}{2}\frac{1}{2}^{-}$ system
are shown in Fig. 2, which shows a multi-resonance behavior. There are three resonance states in the $N\eta_{c}$
scattering phase shifts, corresponding to $\Sigma_{c}D$, $\Sigma_{c}D^{*}$ and $\Sigma^{*}_{c}D^{*}$ states;
while in other scattering channels, there are only two resonance states, which are $\Sigma_{c}D$ and
$\Sigma_{c}D^{*}$. There is only a wavy motion around the threshold of the third state, $\Sigma^{*}_{c}D^{*}$.
The reason is that the channel coupling pushes the higher state above the threshold. The resonance mass and
decay width of resonance states by four-channel coupling are also listed in Table~\ref{mass_c1}.
Both $\Sigma_{c}D$ and $\Sigma_{c}D^{*}$ with $IJ^{P}=\frac{1}{2}\frac{1}{2}^{-}$ are resonance states in
related scattering channels. The resonance mass range of $\Sigma_{c}D$ state is $4306.7
\sim 4311.3$ MeV and the decay width is about $7.1$ MeV, and $\Sigma_{c}D^{*}$ has the mass range of
$4448.8 \sim 4461.6$ MeV and the decay width of $6.2$ MeV. $\Sigma^{*}_{c}D^{*}$ appears as a resonance
state only in the $N\eta_{c}$ channel, with mass of $4525.8$ MeV and decay width of $4.0$ MeV. These results
are qualitatively similar to the conclusion of Ref. \cite{WuJJ}, in which they
predicted two new $N^{*}$ states (the $\Sigma_{c}D$ molecular state $N^{*}(4265)$ and the
$\Sigma_{c}D^{*}$ molecular state $N^{*}(4415)$) in the coupled-channel unitary approach.
\begin{figure*}
\begin{center}
\epsfxsize=5.5in \epsfbox{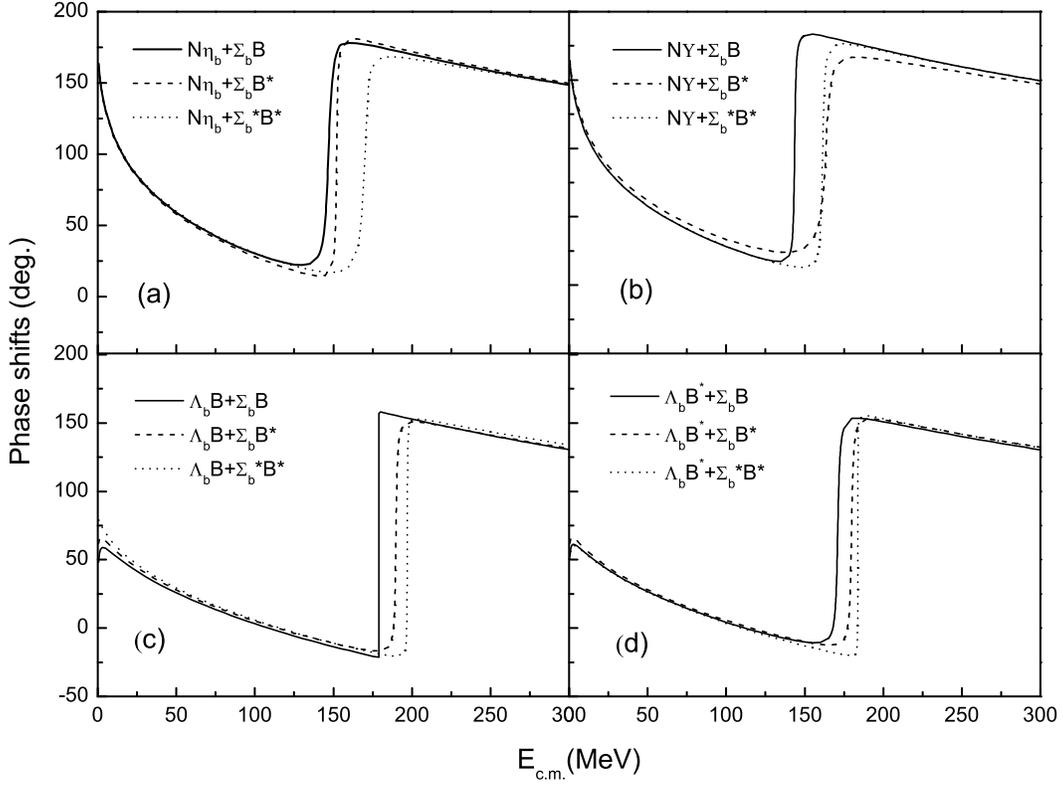} \vspace{-0.1in}

\caption{The $N\eta_{b}$, $N\Upsilon$, $\Lambda_{b}B$ and $\Lambda_{b}B^{*}$ $S-$wave phase shifts
  with two-channel coupling for the $IJ^{P}=\frac{1}{2}\frac{1}{2}^{-}$ system.}
\end{center}
\end{figure*}

\begin{table*}
\begin{center}
\caption{The mass and decay width (in MeV) of the $IJ^{P}=\frac{1}{2}\frac{1}{2}^{-}$ resonance states
  in the $N\eta_{b}$, $N\Upsilon$, $\Lambda_{b}B$ and $\Lambda_{b}B^{*}$ $S-$wave scattering process.}
{\begin{tabular}{@{}c|cccccc|cccccc} \hline
 \multicolumn{1}{c|}{}
 &\multicolumn{6}{c}{two-channel coupling}  & \multicolumn{6}{|c}{four-channel coupling}  \\\hline
\multicolumn{1}{c|}{}
 &\multicolumn{2}{c}{$\Sigma_{b} B$}&\multicolumn{2}{c}{$\Sigma_{b} B^{*}$}&\multicolumn{2}{c|}{$\Sigma_{b}^{*} B^{*}$} &\multicolumn{2}{c}{$\Sigma_{b} B$}&\multicolumn{2}{c}{$\Sigma_{b} B^{*}$}&\multicolumn{2}{c}{$\Sigma_{b}^{*} B^{*}$}\\
 &~~~~~$M$~~~ & ~~~$\Gamma$~~~ & ~~~$M$~~~ & ~~~$\Gamma$~~~ & ~~~$M$~~~ & ~~~$\Gamma$~~~  & ~~~~~$M$~~~ & ~~~$\Gamma$~~~ & ~~~$M$~~~ & ~~~$\Gamma$~~~ & ~~~$M$~~~ & ~~~$\Gamma$~~~ \\
 ~~~{$N\eta_{b}$}~~~        & ~~11083.3 & 4.0 & 11123.9 & 1.4 & 11154.5 & 4.7 & ~~11079.8 & 1.2 & 11120.6 & 0.4 & 11156.9 & 2.0  \\
 {$N\Upsilon$}          & ~~11080.4 & 1.4 & 11135.4 & 6.6 & 11146.2 & 2.0 & ~~11077.5 & 0.1 & 11125.8 & 0.8 & 11153.5     & 3.0   \\
 {$\Lambda_{b}B$}     & ~~11079.0 & 0.0003 & 11125.4 & 2.0 & 11145.1 & 0.49 & ~~11077.2 & 0.001 & 11122.0 & 0.6 & 11141.8 & 0.1 \\
 {$\Lambda_{b}B^{*}$} & ~~11082.25 & 2.6 & 11126.2 & 2.3 & 11142.7 & 0.22 & ~~11078.3 & 0.3 & 11123.0 & 1.2 & 11141.5 & 0.4  \\
  \hline
\end{tabular}
\label{mass_b1}}
\end{center}
\end{table*}
Particularly, in Figs. 1 and 2, the low-energy scattering phase shifts of both $N\eta_{c}$ and $NJ/\psi$
channels go to $180^{\circ}$ at $E_{c.m.}\sim 0$ and rapidly decreases as $E_{c.m.}$ increases, which
implies that both $N\eta_{c}$ and $NJ/\psi$ state are bound states with the help of channel-coupling.
Meanwhile, the slope of the low-energy phase shifts (near $E_{c.m.}\sim 0$) of both $\Lambda_{c}D$ and
$\Lambda_{c}D^{*}$ is opposite to that of $N\eta_{c}$ and $NJ/\psi$ channels, which means that neither
$\Lambda_{c}D$ nor $\Lambda_{c}D^{*}$ state is bound state even with channel coupling.
All these results are consistent with our bound-state calculation~\cite{HuangHX}.

For $IJ^{P}=\frac{1}{2}\frac{3}{2}^{-}$ system, the same calculation has been done and similar results
are obtained. In the two-channel coupling calculation, three bound states $\Sigma_{c}D^{*}$,
$\Sigma^{*}_{c}D$ and $\Sigma^{*}_{c}D^{*}$ all appear as resonance states in the scattering phase
shifts $N J/\psi$ and $\Lambda_{c}D^{*}$, which are shown in Fig. 3. In the four-channel coupling
calculation, the multi-resonance behavior appears again, as shown in Fig. 4. Three resonance states
appear in the $\Lambda_{c}D^{*}$ scattering phase shifts, corresponding to $\Sigma_{c}D^{*}$,
$\Sigma^{*}_{c}D$ and $\Sigma^{*}_{c}D^{*}$ states; while in $N J/\psi$ scattering channels, there are
only two resonance states $\Sigma_{c}D^{*}$ and $\Sigma^{*}_{c}D$. The cusp in the dashed line of
Fig. 4 is a remnant of the $\Sigma^{*}_{c}D^{*}$. The resonance mass and decay width of resonance
states by two kinds of channel coupling are listed in Table~\ref{mass_c2}. The $\Sigma_{c}D^{*}$ is
showed as a resonance state in both $N J/\psi$ and $\Lambda_{c}D^{*}$ scattering process with mass
range of $4452.0 \sim 4454.0$ MeV and decay width of $1.8$ MeV; the $\Sigma^{*}_{c}D$ is also a
resonance state in both $N J/\psi$ and $\Lambda_{c}D^{*}$ scattering channels with mass range of
$4374.4 \sim 4376.4$ MeV and decay width of $2.4$ MeV; while $\Sigma^{*}_{c}D^{*}$ appears as a
resonance state only in the $\Lambda_{c}D^{*}$ channel, with mass of $4523.0$ MeV and decay width
of $1.0$ MeV. It is obvious that the mass of this resonance state $\Sigma^{*}_{c}D$ is consistent
with the $Pc(4380)$, but the decay width is much smaller than the experimental data, which is about
$200$ MeV. As mentioned above, only the hidden-charm channels are considered in this work, so
the total decay width of this state is the lower limits here. More decay channels should be
considered in future work.
\begin{center}
\begin{figure}
\epsfxsize=3.3in \epsfbox{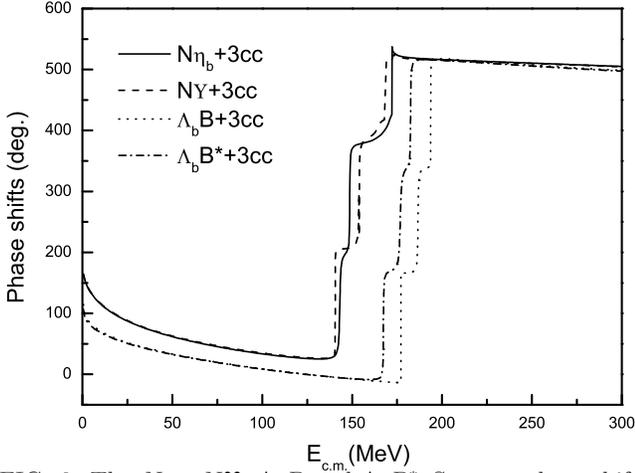} \vspace{-0.2in}

\caption{The $N\eta_{b}$, $N\Upsilon$, $\Lambda_{b}B$ and $\Lambda_{b}B^{*}$ $S-$wave phase shifts
with four-channel coupling for the $IJ^{P}=\frac{1}{2}\frac{1}{2}^{-}$ system.}
\end{figure}
\end{center}

\begin{center}
\begin{figure}
\epsfxsize=3.3in \epsfbox{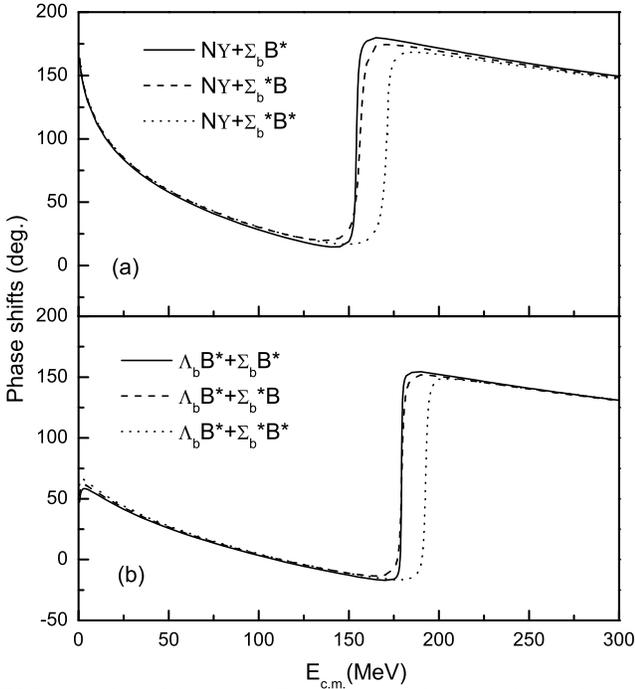} \vspace{-0.2in}

\caption{The $N\Upsilon$ and $\Lambda_{b}B^{*}$ $S-$wave phase shifts with two-channel coupling
for the $IJ^{P}=\frac{1}{2}\frac{3}{2}^{-}$ system.}
\end{figure}
\end{center}

Moreover, the behaviour of the low-energy phase shifts of $NJ/\psi$ channel in both Figs. 3 and 4
is similar to that in Figs. 1 and 2. This indicates that the $NJ/\psi$ with
$IJ^{P}=\frac{1}{2}\frac{3}{2}^{-}$ is possible to be bound by channel-coupling calculation.
By contrast, the slope of the low-energy phase shifts of $\Lambda_{c}D^{*}$ is opposite to that of
$NJ/\psi$ channel, which means that the $\Lambda_{c}D^{*}$ is unbound even with channel coupling.
All these results are also consistent with our previous bound-state calculation in Ref.~\cite{HuangHX}.

Because of the heavy flavor symmetry, we also extend the study to the hidden-bottom pentaquarks.
The results are similar to the hidden-charm molecular pentaquarks.
From Figs. 5 and 6, we can see that the $\Sigma_{b}B$, $\Sigma_{b}B^{*}$ and $\Sigma^{*}_{b}B^{*}$ states
with $IJ^{P}=\frac{1}{2}\frac{1}{2}^{-}$ appear as resonance states in all scattering channels
($N\eta_{b}$, $N\Upsilon$, $\Lambda_{b}B$ and $\Lambda_{b}B^{*}$). The mass and decay width are illustrated
in Table~\ref{mass_b1}. The resonance mass range of $\Sigma_{b}B$ state is $11077.2
\sim 11079.8$ MeV and the decay width is about $1.6$ MeV; $\Sigma_{b}B^{*}$ has the mass range of $11120.6
\sim 11125.8$ MeV and the decay width of $3.0$ MeV; and $\Sigma^{*}_{b}B^{*}$ has the mass range of $11141.5
\sim 11156.9$ MeV and the decay width of $5.6$ MeV. These results are qualitatively similar to the conclusion
of Ref.~\cite{Zou3}, in which they predicted a few narrow $N^{*}$ resonances with hidden beauty around $11$ GeV
in the coupled-channel unitary approach.

For the hidden-bottom pentaquarks with $IJ^{P}=\frac{1}{2}\frac{3}{2}^{-}$, both the states
$\Sigma_{b}B^{*}$ and $\Sigma^{*}_{b}B$ appear as resonance states in the scattering phase shifts of
$N \Upsilon$ and $\Lambda_{b}B^{*}$ channels. The $\Sigma^{*}_{b}B^{*}$ state appear as a resonance state
only in the $\Lambda_{b}B^{*}$ scattering process. All the phase shifts are shown in Figs. 7 and 8.
The mass and decay width are listed in Table~\ref{mass_b2}, from which we can see that the
$\Sigma_{b}B^{*}$ has the mass range of $11122.2 \sim 11122.7$ MeV and the decay width of $0.4$ MeV;
and $\Sigma^{*}_{b}B$ has the mass range of $11102.4 \sim 11103.6$ MeV and the decay width of $1.1$ MeV.
The resonance mass of $\Sigma^{*}_{b}B^{*}$ is $11150.0$ MeV and the decay width is $1.8$ MeV.

Similarly, the behaviour of the low-energy phase shifts of both $N\eta_{b}$ and $N\Upsilon$ also implies
that $N\eta_{b}$ and $N\Upsilon$ with $IJ^{P}=\frac{1}{2}\frac{1}{2}^{-}$, as well as $N\Upsilon$ with
$IJ^{P}=\frac{1}{2}\frac{3}{2}^{-}$ are all possible to be bound with channel-coupling.

\begin{table}
\begin{center}
\caption{The mass and decay width (in MeV) of the $IJ^{P}=\frac{1}{2}\frac{3}{2}^{-}$ resonance states in the $N\Upsilon$ and $\Lambda_{b}B^{*}$ $S-$wave scattering process.}
{\begin{tabular}{@{}c|cccccc} \hline
 & \multicolumn{6}{c}{two-channel coupling}    \\ \hline
 & \multicolumn{2}{c}{$\Sigma_{b} B^{*}$}&\multicolumn{2}{c}{$\Sigma^{*}_{b} B$}&\multicolumn{2}{c}{$\Sigma_{b}^{*} B^{*}$} \\
 & ~~~~~$M$~~~ & ~~~$\Gamma$~~~ & ~~~$M$~~~ & ~~~$\Gamma$~~~ & ~~~$M$~~~ & ~~~$\Gamma$~~~  \\
 ~~~{$N\Upsilon$}~~~   & ~~11126.3& 1.7 & 11105.8 & 4.4 & 11155.7 & 3.8  \\
 $\Lambda_{b}B^{*}$ & ~~11125.5 & 0.9 & 11103.5 & 2.6 & 11152.0 & 2.7  \\  \hline
 & \multicolumn{6}{|c}{four-channel coupling}  \\ \hline
 & \multicolumn{2}{c}{$\Sigma_{b} B^{*}$}&\multicolumn{2}{c}{$\Sigma^{*}_{b} B$}&\multicolumn{2}{c}{$\Sigma_{b}^{*} B^{*}$} \\
 & ~~~~~$M$~~~ & ~~~$\Gamma$~~~ & ~~~$M$~~~ & ~~~$\Gamma$~~~ & ~~~$M$~~~ & ~~~$\Gamma$~~~   \\
 ~~~{$N\Upsilon$}~~~   & ~~11122.7 & 0.2 & 11103.6 & 0.8 & nr & --  \\
 {$\Lambda_{b}B^{*}$}  & ~~11122.2 & 0.2 & 11102.4 & 0.3 & 11150.0 & 1.8  \\ \hline
\end{tabular}
\label{mass_b2}}
\end{center}
\end{table}

\begin{center}
\begin{figure}
\epsfxsize=3.3in \epsfbox{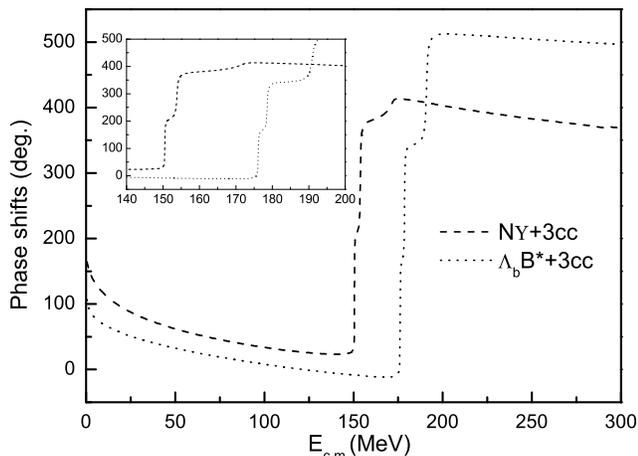} \vspace{-0.2in}

\caption{The $N\Upsilon$ and $\Lambda_{b}B^{*}$ $S-$wave phase shifts with four-channel coupling
for the $IJ^{P}=\frac{1}{2}\frac{3}{2}^{-}$ system.}
\end{figure}
\end{center}

\section{Summary}
In summary, we investigate the hidden-charm and hidden-bottom pentaquark resonances in the hadron-hadron
scattering process. For the hidden-charm sector, three resonance states with
$IJ^{P}=\frac{1}{2}\frac{1}{2}^{-}$, as well as three resonance states with
$IJ^{P}=\frac{1}{2}\frac{3}{2}^{-}$ are found to be dynamically generated from coupled scattering channels.
Because of the hidden $c\bar{c}$ components involved in these states, the masses of these states are all
above $4.2$ GeV while their widths are only a few MeV. Extending to the hidden-bottom system,
the results are similar. Both the resonance states with $IJ^{P}=\frac{1}{2}\frac{1}{2}^{-}$ and
$IJ^{P}=\frac{1}{2}\frac{3}{2}^{-}$ are found from corresponding scattering process. The masses of these
states are all above $11$ GeV while their widths are only a few MeV. The nature of these states is similar
to the corresponding $N^{*}_{c\bar{c}}$ and $N^{*}_{b\bar{b}}$ states predicated in Ref.~\cite{WuJJ} and
Ref.~\cite{Zou3}, which definitely cannot be accommodated by the conventional $3q$ quark models, and should
form part of the heavy island for the quite stable $N^{*}$ baryons.

Particularly, the behaviour of the low-energy phase shifts of the $N\eta_{c}$, $NJ/\psi$, $N\eta_{b}$ and
$N\Upsilon$ indicates that the states $N\eta_{c}$, $NJ/\psi$, $N\eta_{b}$ and $N\Upsilon$ with
$IJ^{P}=\frac{1}{2}\frac{1}{2}^{-}$, as well as $NJ/\psi$ and $N\Upsilon$ with
$IJ^{P}=\frac{1}{2}\frac{3}{2}^{-}$ are all possible to be bound by channel-coupling calculation.

All these heavy pentaquarks are worth searching in future experiments. Immediately after the LHCb,
the Jefferson Lab proposed to look for the hidden-charm pentaquarks by using photo-production of
$J/\psi$ at threshold in Hall C~\cite{Jlab1}. Moreover, the pentaquarks with charm quarks can also
be observed by the PANDA/FAIR~\cite{Panda}. For the pentaquarks with the hidden-bottom, we hope
the proposed electron-ion collider (EIC)~\cite{EIC} and the Jefferson Lab~\cite{Jlab2} to discover
these interesting super-heavy pentaquarks.

\section*{Acknowledgment}
This work is supported partly by the National Science Foundation
of China under Contract Nos. 11675080, 11775118 and 11535005, the Natural Science Foundation of
the Jiangsu Higher Education Institutions of China (Grant No. 16KJB140006).

\end{document}